# Electronic and Lattice Dynamical Properties of $Ti_2SiB$ MAX Phase

Aysenur Gencer[a] and Gokhan Surucu[a,b,c,*]

[a]*Department of Physics, Middle East Technical University, Ankara, Turkey;* [b]*Department of Electric and Energy, Ahi Evran University, Kirsehir, Turkey;* [c]*Photonics Application and Research Center, Gazi University, Ankara, Turkey*

*Corresponding author. Tel.: +90 312 210 4322; E-mail address: g_surucu@yahoo.com (Gokhan Surucu)
Postal address: METU Physics Department Universiteler Mah. Dumlupinar Bul. No:1 Cankaya 06800 Ankara Turkey*

**Abstract**

The structural, electronic, mechanic, vibrational and thermodynamic properties of $Ti_2SiB$ which is a hypothetical MAX phase compound, have been investigated using density functional theory calculations. The structural optimization of $Ti_2SiB$ has been performed and the results have been compared with $Ti_2SiC$, $Ti_2SiN$, and $Ti_2AlB$ that are studied in the literature. Then the band structure and corresponding partial density of states are computed. In addition, charge density and Bader charge analysis have been performed. The elastic constants have been obtained, then the secondary results such as bulk modulus, shear modulus, Young's modulus, Poisson's ratio, and Vickers Hardness of polycrystalline aggregates have been derived, and the relevant mechanical properties have been discussed. Moreover, the elastic anisotropy has been visualized in detail by plotting the directional dependence of compressibility, Poisson ratio, Young's and Shear moduli. Furthermore, the phonon dispersion curves as well as corresponding phonon PDOS, and thermodynamical properties such as free energy, entropy and heat capacity have been computed and the obtained results have been discussed in detail. This study provides the first considerations of $Ti_2SiB$ that could have a potential application in nuclear industry.

# 1. Introduction

MAX phases have $M_{(n+1)}AX_n$ formula with n=1,2 and 3 where M is a transition metal, A is an A group element and X is Carbon and/or Nitrogen [1]. The MAX phases crystallize in $P6_3/mmc$ hexagonal structure and their ionic, metallic and covalent bonding give them unique properties [2]. The MAX phases have both metallic and ceramic properties that lead to good damage tolerance, high strength and stiffness at high temperatures, good electrical and thermal conductivity and good corrosion resistance [3–5]. Owing to these properties, the MAX phases can be implemented for applications such as wear and corrosion resistant coatings [6], superconducting materials [7], and nuclear industry [8].

The MAX phases are studied experimentally, for example, Hu et al. [9] fabricated *Nb$_4$AlC$_3$* using spark plasma sintering and investigated its thermal expansion and electrical conductivity and Barsoum and Radorovic [5] reviewed the results of the measurement of elastic properties of some selected MAX phases. Also, theoretical study of MAX phases are available in the literature, for example, Bouhemadou and Khenata [10] studied the structural, elastic and electronic properties of *M$_2$SC* (*M=Ti, Zr, Hf*) and He et al. [11] studied the lattice constants, Bulk modulus, band structure and partial density of states for *Ti$_2$InC*, *Zr$_2$InC*, and *Hf$_2$InC*. The given studies are examples and there are many studies for MAX phases. Furthermore, studies of the MAX phases are focused on different combinations of M- and A- site atoms and X site atom is kept C and/or N [2]. Recently, MXenes which are 2D materials, are layered MAX phases and has been investigated to explore their properties[12]. Moreover, the MAX phases of *Ti$_3$SiC$_2$*, *Ti$_3$AlC$_2$* and *Ti$_2$AlC* have been examined how they behave under neutron radiation [13] and they are radiation hard similar to *SiC* which is one of the most used material in nuclear reactors. On the other hand, Borides of MAX phases would be an alternative new material for control rods at nuclear reactors due to having the high neutron cross section of Boron. Moreover, Borides of MAX phases would be used in nuclear industry due to their stability. There are very few studies that have X site as Boron [14–16] and therefore, the aim of this study is to determine the properties of *Ti$_2$SiB* compound which has Boron for X site atom and could have potential application in nuclear industry.

In this study, *Ti$_2$SiB* is investigated for the first time as we know up to date and compared with *Ti$_2$SiC* and *Ti$_2$SiN* that are detailed studied in the literature [17–27] and also *Ti$_2$AlB* from a previous study [15]. Calculation details for *Ti$_2$SiB* are illustrated in Section 2. The structural optimization results, electronic properties and charge density analysis of *Ti$_2$SiB* are presented in Section 3. The calculated formation enthalpy indicates that *Ti$_2$SiB* is a stable phase. After these results, mechanical properties such as elastic constants, bulk modulus, shear modulus, etc. are detailed in Section 4. The phonon band structure and thermodynamic properties are presented in Section 5 where the Raman modes are also included. Finally, the study concludes with a brief summary in Section 6.

## 2. Calculation Details

The DFT calculations are performed using the Vienna Ab initio Simulation Package (VASP) [28]. The pseudopotentials are chosen according to Perdew-Burke-Ernzerhof (PBE) parametrization of the generalized gradient approximation (GGA) for the exchange and correlation terms of the electron-electron interaction [29]. For the electron-ion interaction, Projector Augmented Wave (PAW) method [30] has been implemented and the kinetic energy cut off is chosen 500 eV. The 16 x 16 x 4 k-point mesh has been generated which is centered at the Γ- point. The electronic energy change is kept $10^{-11}$ eV. The structure is optimized with the minimization of the stresses and the Hellman-Feynman forces. The elastic constants are calculated using the stress-strain method with VASP. The anisotropic elastic properties are obtained with ELATE program [31] where the calculated elastic constants are employed. The phonon calculations are performed for a 2 x 2 x 1 supercell with the Phonopy code [32].

## 3. Structural Optimization and Electronic Properties

The lattice structure of $Ti_2SiB$ compound which belongs to the space group of 194 (P6$_3$/mmc) is given in Figure 1. The $Ti$ atoms are interleaved with the $Si$ atoms and the octahedral sides are filled with the $B$ atoms as shown in Figure 1. The $Ti$ atoms occupy at 4f, the $Si$ atoms occupy at 2d and the $B$ atoms occupy at 2a Wyckoff positions.

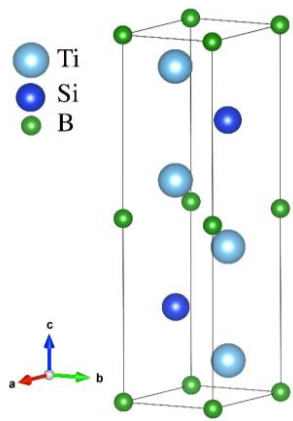

Figure 1. The lattice structure of Ti$_2$SiB compound

Lattice parameters, formation energy and z parameter of the Wyckoff positions of the *Ti$_2$SiB* compound are given in Table 1 and the results from the literature for *Ti$_2$SiC* [22,24], *Ti$_2$SiN* [17,26] and *Ti$_2$AlB* [15] are also listed. The lattice parameters are increasing when the *X* atom goes from nitrogen to boron for *Ti$_2$SiX* as can be seen in Table 1. Once *Ti$_2$SiB* and *Ti$_2$AlB* are compared, c parameter of *Ti$_2$AlB* is around 8% higher than that of *Ti$_2$SiB*. The *Ti$_2$SiB* compound is hypothetical therefore there is no experimental data nor theoretical study in order to compare the calculation results. The formation energy is calculated using Equation 1 and negative formation energy indicates that the compound is thermodynamically stable. Hence, the results specify that these compounds can be synthesized.

$$E_{Formation} = E_{Ti_2SiB} - 2E_{Ti} - E_{Si} - E_B \qquad (1)$$

Table 1. Lattice parameters, formation energy and z parameter of the Wyckoff positions of the Ti$_2$SiB, Ti$_2$SiC and Ti$_2$SiN compounds

| Material | Reference | a (Å) | c (Å) | ΔH$_f$ (eV/atom) | z |
|---|---|---|---|---|---|
| Ti$_2$SiB | This study | 3.151 | 12.979 | -3.972 | 0.095 |
| Ti$_2$SiC | Theory [22] | 3.052 | 12.873 | - | 0.092 |
| Ti$_2$SiC | Theory [24] | 3.052 | 12,873 | -0.860 | 0.092 |
| Ti$_2$SiN | Theory [17] | 2.979 | 12.82 | - | 0.093 |
| Ti$_2$SiN | Theory [26] | 2.984 | 12.822 | - | 0.093 |
| Ti$_2$AlB | Theory [15] | 3.148 | 14.064 | -3.577 | 0.087 |

The band structure for *Ti$_2$SiB* is predicted along the high symmetry directions in the first Brillouin zone from the calculated equilibrium lattice constant. The band structures and corresponding partial and total electronic density of state (DOS) are displayed drawn and given in Figure 2 and Figure 3, respectively.

It is clear that the compound is metallic due to the fact that the DOS values differ from zero at the Fermi level.

The most significant contribution to PDOS comes from the s states of *B* and *Si* atoms between -7 eV and -5 eV energy range and from d states of *Ti* between -2 eV and 3 eV energy range as seen in Figure 3.

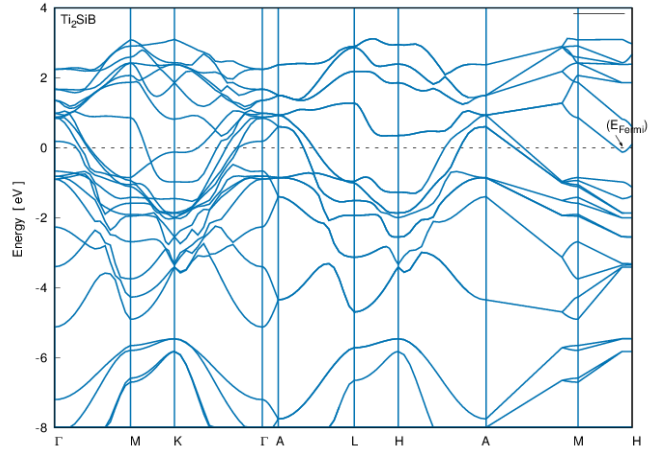

Figure 2. Band structure of $Ti_2SiB$

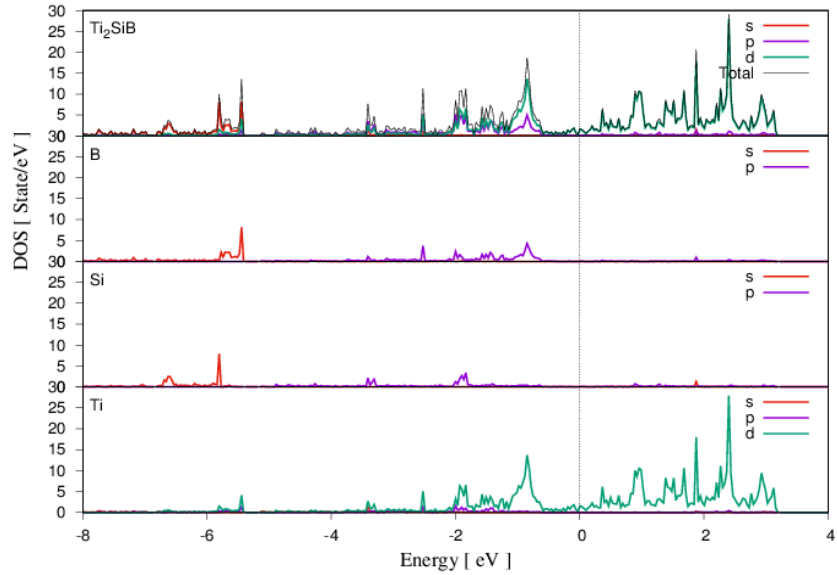

Figure 3. Partial density of states of $Ti_2SiB$

After these electronic structure calculations, the charge density and bader charge analysis is performed in order to determine the bonding nature and charge of the ions for *Ti₂SiB*. The charge density plot of *Ti₂SiB* indicates that *Ti₂SiB* has dominantly ionic bonding as shown in Figure 4. Bader charge population analysis is also performed in order to get the bonding nature of *Ti₂SiB*. The calculation is performed with VASP and analysis is performed using the algorithm developed by Henkelman et. al. [33] which is based on Bader's suggestion [34]. Table 2 lists the Bader net charge of the ions for *Ti$_2$SiB*. If the Bader net charge is positive, the charge is transferred away from the atom, vice versa for the negative Bader net charge [35]. The charge is transferred away from *Ti* atom, while the charge is transferred to *Si* and *B* atoms. The bond length of *Ti-Si* is 2.712 Å and it is 2.198 Å for *Ti-B*.

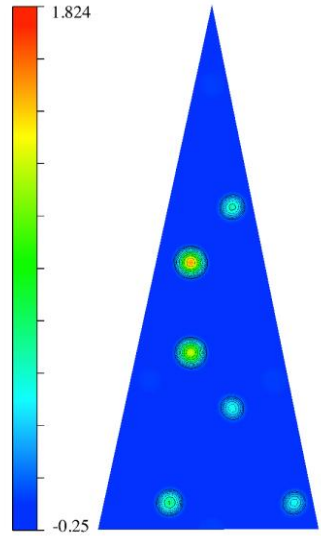

Figure 4. The charge density of Ti$_2$SiB compound

Table 2: Bader net charge of the atoms for Ti$_2$SiB in units of e

| Atom | Bader Charge |
|------|--------------|
| Ti   | 0.552        |
| Si   | -0.282       |
| B    | -0.822       |

**4. Mechanical Properties**

Mechanical properties of *Ti$_2$SiB* are investigated after the structural optimization. The elastic constants (C$_{ij}$) are calculated using the stress-strain method with VASP which are listed in Table 3. In addition, to compare the obtained results with the literature, the table contains the results for *Ti$_2$SiC* [22,24], *Ti$_2$SiN* [17,26], and *Ti$_2$AlB* [15]. Also, the elastic constants of *Ti$_2$SiB* has lower value when compared to the values of *Ti$_2$SiC and Ti$_2$SiN*. On the other hand, this value is higher than the value given for *Ti$_2$AlB*. The stability of *Ti$_2$SiB* are determined using Born stability criteria [36] and found that it is mechanically stable.

Table 3: The calculated elastic constants (in GPA) of the $Ti_2SiB$ and results for $Ti_2SiC$ and $Ti_2SiN$ taken from the literature

| Material | Reference | $C_{11}$ | $C_{12}$ | $C_{13}$ | $C_{33}$ | $C_{44}$ |
|---|---|---|---|---|---|---|
| $Ti_2SiB$ | This study | 250.1 | 74.8 | 80.9 | 262.8 | 119.6 |
| $Ti_2SiC$ | Theory [23] | 311.4 | 85.8 | 111.5 | 324.2 | 146.1 |
| $Ti_2SiC$ | Theory [19] | 311.0 | 84.0 | 107.0 | 343.0 | 155.0 |
| $Ti_2SiN$ | Theory [17] | 298.0 | 96.0 | 127.0 | 347.0 | 153.0 |
| $Ti_2SiN$ | Theory [26] | 296.7 | 100.2 | 126.3 | 347.8 | 155.1 |
| $Ti_2AlB$ | Theory [15] | 234.0 | 73.9 | 80.6 | 261.9 | 115.1 |

The mechanical properties of *Ti₂SiB*, which are obtained with the calculation of the elastic constants, are listed in Table 4. The literature results for *Ti₂SiC* and *Ti₂SiN* are also given in Table 4. The Bulk modulus (B) and the shear modulus (G) are calculated with the Voigt-Reuss-Hill approximations [37–39]. Voight approximation gives the lower limit while Reuss approximation gives the upper limit of these moduli. Hill approximation takes the average of the Voigt and Reuss results which is generally consistent with the experimental results. As it is well known that Bulk modulus (B) is the volume change of a material if there is a stress on it. So, it defines the incompressibility of a material. The Bulk modulus of *Ti₂SiX* compound increases when *X* element goes from Boron to Nitrogen. On the other hand, if *Ti₂SiB* and *Ti₂AlB* are compared, *Ti₂AlB* has higher Bulk modulus then *Ti₂SiB*. Shear modulus (G) is also known as the length change of a material when a shear stress is applied. So, it is a measure of the resistance of the transverse deformations. Once the Shear modulus of *Ti₂SiB* compound is compared, it has lower Bulk modulus than that of *Ti₂SiC and Ti₂SiN* while it has higher Bulk modulus than *Ti₂AlB*. Young's modulus (E) is defined as the length change of a material when a pull or push is applied. It is also called modulus of elasticity. Young's modulus of *Ti₂SiB* shows similar behavior with the Shear modulus. Poisson's ratio (ϑ) is the ratio of the transverse strain to axial strain. It is used to determine the bonding characteristics of a material. The material has ionic bonding if ϑ is around 0.25 while if ϑ is small around 0.1, it has covalent bonding [40]. ϑ value of *Ti₂SiB* compound indicates that it has dominantly ionic bonding. Also, the literature results for *Ti₂SiC, Ti₂SiN and Ti₂AlB* show that they both have dominantly ionic bonding. B/G ratio determines the ductility or brittleness of a material and if it is higher than 1.75, the material is ductility, otherwise, it is brittle [41,42]. *Ti₂SiB* compound is brittle that can be inferred from B/G ratio of it. G/B ratio is called Pugh's modulus and it is used to determine the bonding nature of a material. If G/B is around 1.1, the material has dominantly covalent bonding [40]. On the other hand, the ionic character is dominant if G/B ratio is around 0.6 [40]. *Ti₂SiB* compound has dominantly ionic bonding as can be concluded from G/B ratio which is consistent with the results from ϑ value and the charge density analysis. The hardness of *Ti₂SiB* is also calculated using Chen et. al. approach [43]. As shown in Table 4, *Ti₂SiB* has the hardness (17.6 GPa) value that is higher than the value given for *Ti₂AlB (16.3 GPa)*. Thus, *Ti₂SiB* could be considered as the hard material.

Table 4: The calculated mechanical properties (Bulk Modulus – B (in GPa), Shear modulus – G (in GPa), Young's modulus - E (in GPa), Poisson's ratio – ϑ, B/G ratio, G/B ratio, Vicker's hardness - $H_v$ (in GPa)) of Ti₂SiB and Ti₂SiC and Ti₂SiN taken from literature.

| Material | Reference | B | G | E | ϑ | B/G | G/B | $H_v$ |
|---|---|---|---|---|---|---|---|---|
| Ti₂SiB | This study | 137.2 | 99.3 | 239.9 | 0.208 | 1.381 | 0.723 | 17.6 |
| Ti₂SiC | Theory [23] | 173.6 | 122.0 | 296.6 | 0.215 | - | 0.70 | - |
| Ti₂SiC | Theory [19] | 172.5 | 128.5 | 309 | 0.20 | 1.34 | - | - |
| Ti₂SiN | Theory [17] | 182 | 118 | 291 | 0.233 | - | - | - |
| Ti₂SiN | Theory [26] | 181.9 | 117.8 | 290.7 | 0.234 | - | - | - |
| Ti₂AlB | Theory [15] | 134.0 | 93.6 | 227.4 | 0.215 | 1.420 | 0.703 | 16.3 |

The anisotropic elastic properties are studied because this calculation is important in order to complete the elastic properties of a material. The anisotropic properties provide to determine some material properties under a plastic deformation or dislocation dynamics. The anisotropy is higher if the spherical shape is distorted. The elastic stiffness matrix is calculated using VASP. This matrix is employed to ELATE program [31]. Figure 5 shows Young's modulus, linear compressibility, Shear modulus and Poisson's ratio in xy, xz and yz planes. The green curves show the minimum and the blue curves show the maximum points for the parameters. The maximum and minimum values for these parameters are given in Table 5. Furthermore, the results for $Ti_2AlB$ [15] are also listed in Table 5 as well. Young's modulus and linear compressibility are isotropic in all planes. Shear modulus and Poisson's ratio are anisotropic in xy and xz planes, while they are isotropic in yz plane.

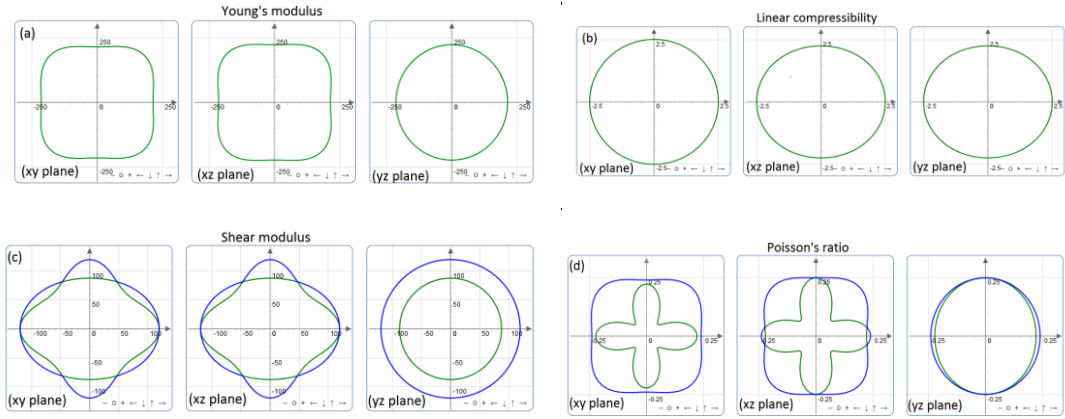

Figure 5. Young's modulus (a), linear compressibility (b), Shear modulus (c) and Poisson's ratio (d) in xy, xz and yz planes for $Ti_2SiB$

Table 5: The maximum and minimum values of Young's modulus (E, in GPa), linear compressibility (β), shear modulus (G, in GPa) and Poisson's ratio (ϑ) of $Ti_2SiB$

|  | Reference | Young's modulus | | Linear compressibility | | Shear Modulus | | Poisson's ratio | |
| --- | --- | --- | --- | --- | --- | --- | --- | --- | --- |
|  |  | $E_{min}$ | $E_{max}$ | $\beta_{min}$ | $\beta_{max}$ | $G_{min}$ | $G_{max}$ | $\vartheta_{min}$ | $\vartheta_{max}$ |
| $Ti_2SiB$ | This study | 214.11 | 261.87 | 2.26 | 2.52 | 87.62 | 119.63 | 0.07 | 0.29 |
| $Ti_2AlB$ | Theory [15] | 197.67 | 252.56 | 2.17 | 2.68 | 1.00 | 115.10 | 0.09 | 0.30 |

## 5. Vibrational and Thermal Properties

The dynamical stability of $Ti_2SiB$ is calculated using finite displacement method in the Phonopy code for a 2x2x1 supercell. The force constants and phonon dispersion frequencies are obtained. The phonon dispersion curve is shown in Figure 6 with partial density of states. $Ti_2SiB$ is a dynamically stable material which has only real phonon branches. There are 24 branches where three of them is acoustic and remaining of them are optical. For these branches, *Ti* atoms contribute at the lower frequencies and *B* atoms contribute higher frequencies as can be seen in Figure 6.

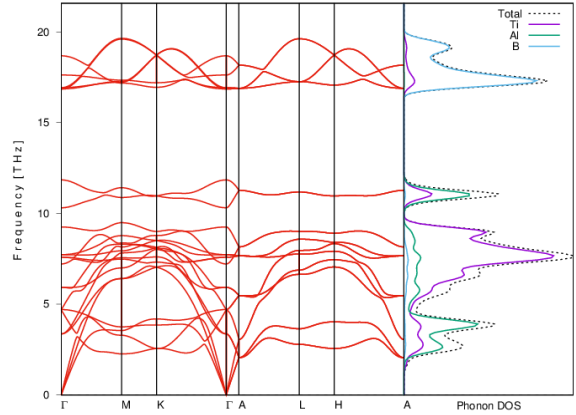

Figure 6. Phonon band structure and partial density of states for Ti$_2$SiB

Additionally, the phonon frequencies at $\Gamma$ point for *Ti$_2$SiB* have been listed in Table 6. The classification of the phonon modes for this material can be given as; A$_{1g}$+3A$_{2u}$+2B$_{1g}$+2B$_{2u}$+2E$_{2u}$+2E$_{2g}$+3E$_{1u}$+E$_{1g}$.

Table 6: The calculated phonon frequencies (THz) at $\Gamma$ point of Ti$_2$SiB.

| Symmetry | |
|---|---|
| E$_{1u}$ | 0.001 |
| A$_{2u}$ | 0.002 |
| E$_{2u}$ | 3.364 |
| B$_{2u}$ | 4.674 |
| E$_{2g}$ | 4.704 |
| E$_{1u}$ | 5.927 |
| B$_{1g}$ | 7.216 |
| E$_{1g}$ | 7.588 |
| E$_{2g}$ | 7.729 |
| A$_{1g}$ | 9.254 |
| A$_{2u}$ | 10.312 |
| B$_{1g}$ | 11.874 |
| E$_{2u}$ | 16.861 |
| $E_{1u}$ | 16.929 |
| $A_{2u}$ | 17.629 |
| $B_{2u}$ | 18.687 |

Here, $A_{2u}+E_{1u}$ belongs to acoustic phonon modes and the others belong to varied optic phonon modes. These optic modes are Raman Active Modes ($\Gamma_R = A_{1g}+E_{1g}+2E_{2g}$), Hyper Raman Active Modes ($\Gamma_H = 2B_{1g}+2B_{2u}+2E_{2u}$) and Infrared Active Modes ($\Gamma_I = 2A_{2u} + 2E_{1u}$). Phonon frequencies of these compounds at $\Gamma$ point can provide useful information for future experiments to identify the predicted new phases.

The thermal properties are calculated after phonon calculations. Figure 7 shows the entropy, free energy and enthalpy as a function of temperature. The entropy increases while the free energy decreases as the temperature goes from 0 to 2000 K as can be seen from Figure 7. The enthalpy also increases linearly with the temperature after 300 K as can be seen from Figure 7. Figure 8 shows the heat capacity as a function of temperature. It is realized from the figure that when T < 750 K, the Cv increases very rapidly with the temperature; when T > 750 K, the Cv increases slowly with the temperature, and it almost approaches a constant value called as Dulong–Petit limit for this compound.

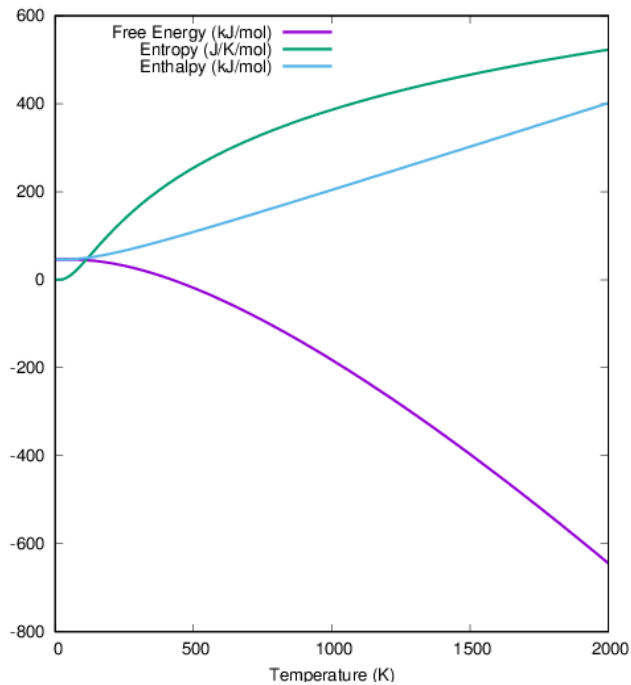

Figure 7. Entropy, free energy and enthalpy as a function of temperature for $Ti_2SiB$

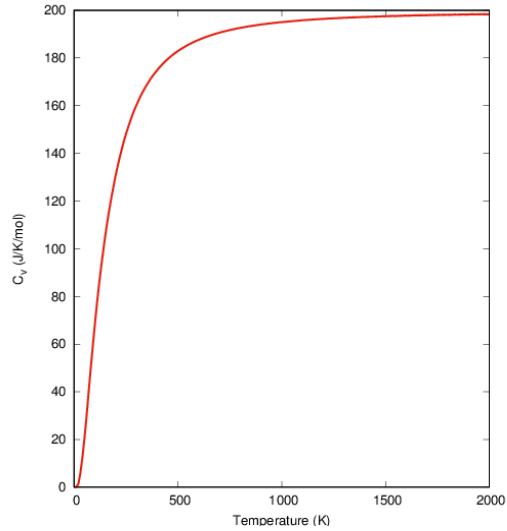

Figure 8. Heat capacity as a function of temperature for $Ti_2SiB$

## 6. Conclusion

The structural, mechanic, electronic, and dynamic properties of *$Ti_2SiB$* have been calculated using VASP. The formation enthalpy of *$Ti_2SiB$* indicates that this compound is energetically stable and therefore could be synthesized. The band structure shows that *$Ti_2SiB$* has a metallic character. Moreover, the charge density illustrates that *$Ti_2SiB$* has dominantly ionic bonding. The calculated elastic constants showed that this compound is mechanically stable. In the calculated phonon dispersion curves, there are no soft modes at any wave vectors, which confirms the dynamical stability of *$Ti_2SiB$*.

Furthermore, the Raman frequencies are obtained in order to offer practical information for the future experiments. The study has been completed with the investigation of thermodynamic properties of *$Ti_2SiB$*. Consequently, in this study, *$Ti_2SiB$* compound, a hypothetical MAX phase, has been investigated and due to its B atom, it is a possible candidate material for the nuclear applications. This is the first study of *$Ti_2SiB$* as best of our knowledge that could provide insights for both theoretical and experimental studies.